\documentstyle[11pt,newpasp,twoside,epsf]{article}
\def\edcomment#1{\iffalse\marginpar{\raggedright\sl#1\/}\else\relax\fi}
\reversemarginpar

\begin{document}
\title{NGC6822: short period variable stars, stellar population and distance scale.}
\author{Lara Baldacci}
\affil{Bologna University, via Ranzani 1, I-40127 Bologna}
\author{Luca Rizzi}
\affil{INAF - Padova Observatory, vicolo dell'Osservatorio 5 I-35122 Padova}
\author{Gisella Clementini}
\affil{INAF - Bologna Observatory, via Ranzani 1 I-40127 Bologna}
\author{Enrico V. Held}
\affil{INAF - Padova Observatory, vicolo dell'Osservatorio 5 I-35122 Padova}
\begin{abstract}
Results are presented on a study of the short period variable stars in the 
dwarf irregular galaxy NGC6822. We observed an almost uniformely populated 
classical instability strip from the Horizontal Branch up to the Classical 
Cepheids region. The main goal we achieved from the analysis of the faint 
sample is the first detection of RR Lyrae stars in this galaxy.  
\end{abstract}

\section{Introduction}
The aim of the present project is the study of the dwarf irregular (dI) galaxy 
NGC6822 using its pulsating variable star content.  
Pulsating variables are tracers of different stellar population in galaxies 
and useful for testing theoretical models. Moreover, some of them, such as the 
RR Lyrae stars and the Classical Cepheids, are primary distance indicators 
in the Local Group.

\section{Observation and analysis}
Time-series imaging of NGC6822 was obtained at the VLT (36 V and 11 B
images) in 2001 August. Photometric reductions were performed using DAOPHOT 
and ALLFRAME 
(Stetson 1994) and the candidate variables were identified using ISIS2.1 
(Alard 2000). About 450 bona fide candidate variables were detected, and all
are being analysed with GRATIS (GRaphycal Analyzer of TIme Series, 
see Clementini et al. 2000) in order to check their actual variation, to 
classify them and to determine their pulsational characteristics.  

\begin{figure}[h]
\plotone{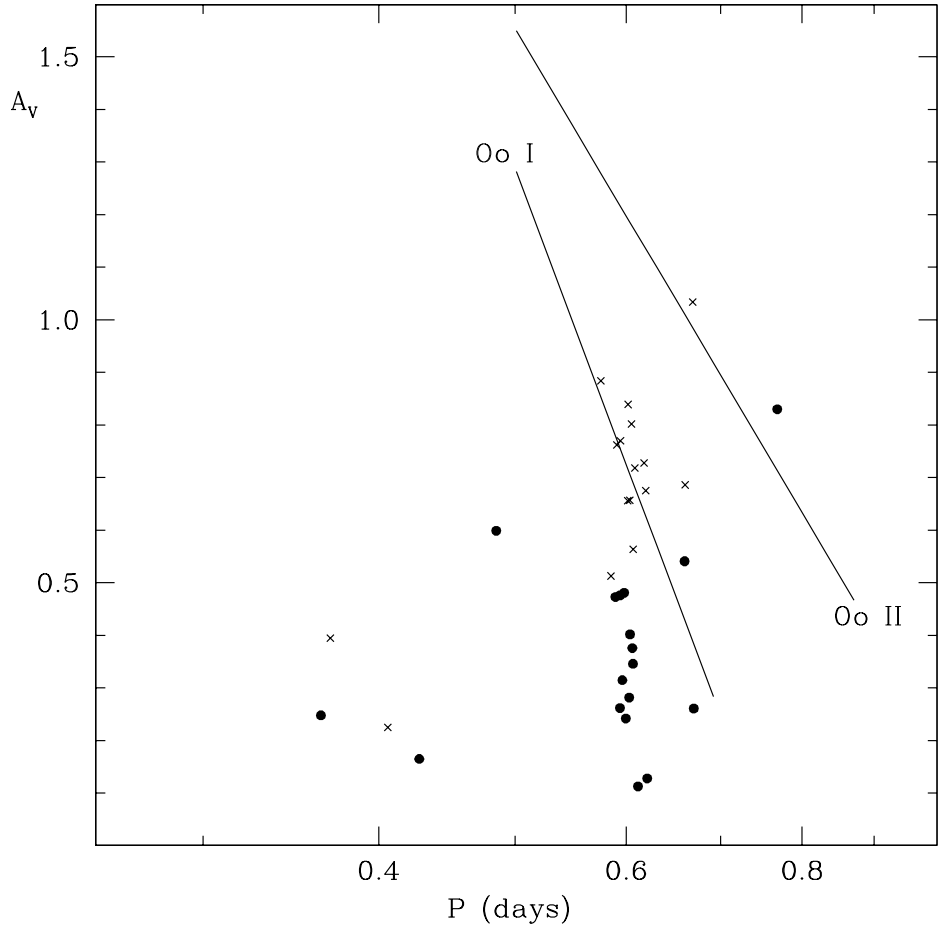}
\caption{P/A relation for RR Lyrae stars (crosses) and low luminosity (LL) 
Cepheids (dots). Lines 
show the P/A relations for Oosterhoff type I and II Galactic globular cluster 
types (Clement \& Rowe 2000).}
\end{figure}
\section{Results}
The color magnitude diagram of NGC6822 shows a significant population 
of candidate variable stars of different types that populate all the 
regions where the theoretical models predict their presence (Baldacci et al. 
2002). 
Only the variables with $V\geq23$ (namely variables in the 
Horizontal Branch region of the classical instability strip) have been fully 
analysed so far. Among them we found  
17 RR Lyrae stars (15 ab-type and 2 c-type) and 20 brighter short-period 
variables that we collectively indicate as low luminosity (LL) Cepheids 
(Clementini et al. 2003). 
All these objects have good photometric data and light curve coverage 
that allow an unambiguous determination of the period. Since  
the Horizontal Branch of NGC6822 in completely hidden by the young and 
intermediate-age star populations, our detection of RR Lyrae stars is the 
first indisputable evidence of the presence of an old stellar population 
($\rm{t\geq10}$ Gyr) in the galaxy.
\begin{figure}[t]
\plotone{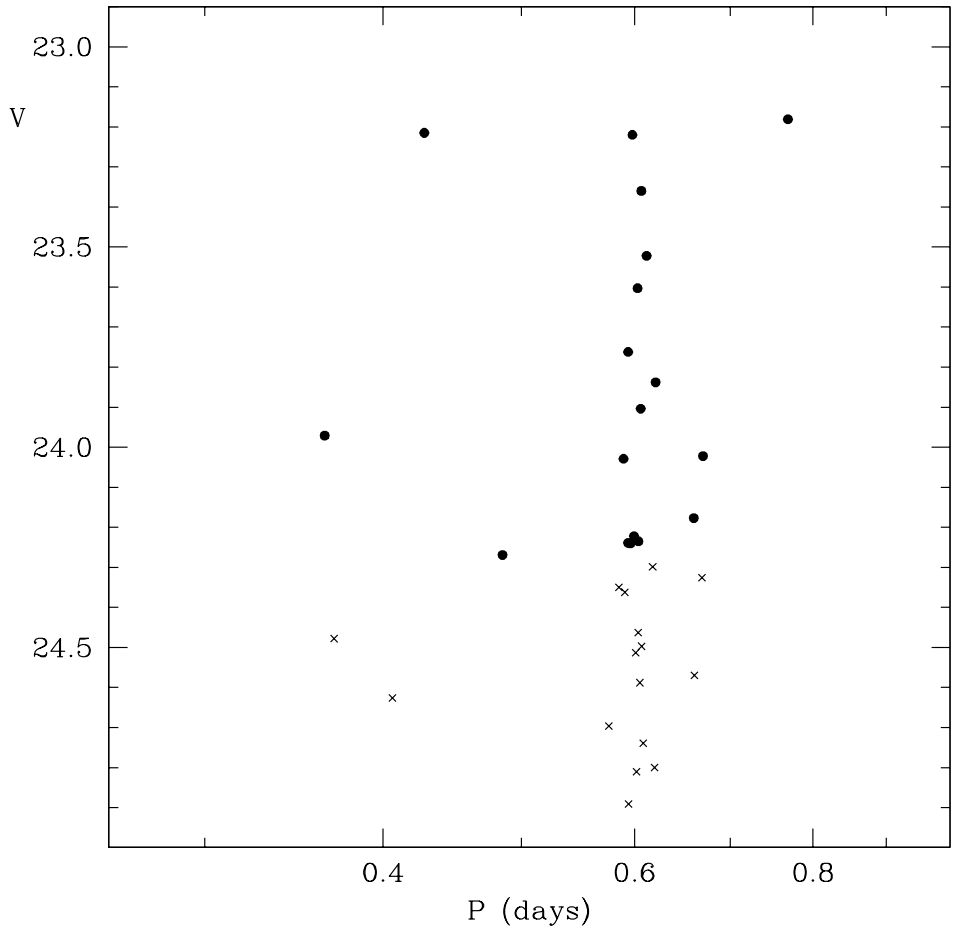}
\caption{P/L relation for RR Lyrae stars (crosses) and low luminosity (LL) 
Cepheids (dots).}
\end{figure}

Figures 1 and 2 show the period-amplitude (P/A) and the period-luminosity
 (P/L) distributions of our sample, respectively. 
The P/A relation of the RR Lyrae stars is similar to the distribution followed
by the Galactic globular clusters of Oosterhoff type I (Oosterhoff 1939), 
although an intermediate distribution (as found in many other dwarf 
galaxies) is not definitely ruled out by our data. The LL Cepheids are characterizated 
by smaller amplitudes (at a given period) than the RR Lyrae stars, and have
similar periods but lower luminosities than the short periods Cepheids 
originally found in the Small Magellanic Cloud by Smith et al. (1992).       	 
The P/L distribution shows that there is no clear gap between 
RR Lyrae stars and LL Cepheids: the latter seem to fill up the region between 
the RR Lyrae stars and the Classical Cepheids in the classical instability 
strip. A similar continuity in the classical instability strip 
has been found in the dI galaxy Phoenix by Gallart et al. 
(2003). The nature of LL Cepheids is still unclear. Because of their 
distribution in the P/L diagram, they could be related both to the short 
periods Cepheids found in other dI galaxies (IC1613: Dolphin et al. 2001;
Leo A: Dolphin et al. 2002; Sextans A: Dolphin et al. 2003; SMC: Sharpee et 
al. 2002) and to the Anomalous Cepheids found in the dwarf spheroidal galaxies 
(e.g., Pritzl et al. 2002 and references therein).  

The mean apparent magnitude of our RR Lyrae stars is 
$V=24.56\pm0.18$ (the error is the standard deviation of the average)
while the mean apparent magnitude of the subsample of variables
located inside the classical instability strip (6 stars) is 
$V=24.63\pm0.14$, yielding a distance modulus of 
$(m-M)_0=23.36\pm0.17$ (see Clementini et al. 2003 for details). 
The LL Cepheids in our faint sample are on average $V\sim0.7$ mag brighter 
than the RR Lyrae. Their period distribution peaks at $P\sim0.6$ d. 
The lack of LL Cepheids with longer periods could be due to selection effects, since our 
observations were optimized to provide good coverage of the light curves for 
variable stars with periods between 0.2 and 0.7 d. We expect 
to find variables with longer periods in the brighter sample we are currently
investigating.

\end{document}